\documentclass[10pt,aps,prl,showpacs,showkeys,twocolumn]{revtex4-1}
\usepackage{amsmath, amssymb}
\usepackage{amsthm}
\usepackage{graphicx}
\usepackage{booktabs}
\usepackage{dcolumn}
\usepackage{textcomp}
\usepackage{multirow}
\usepackage[customcolors]{hf-tikz}

\DeclareMathOperator{\tr}{Tr}

\renewcommand{\vec}[1]{\mathbf{#1}}

\tikzset{style green/.style={
    set fill color=green!50!lime!60,
    set border color=white,
  },
  style cyan/.style={
    set fill color=cyan!90!blue!60,
    set border color=white,
  },
  style orange/.style={
    set fill color=orange!80!red!60,
    set border color=white,
  },
  hor/.style={
    above left offset={-0.15,0.31},
    below right offset={0.15,-0.125},
    #1
  },
  ver/.style={
    above left offset={-0.1,0.3},
    below right offset={0.15,-0.15},
    #1
  }
}

\begin{document}

\title{Recovering quantum properties of continuous-variable states in the presence of measurement errors}
\author{E. Shchukin}
\email{evgeny.shchukin@gmail.com}
\author{P. van Loock}
\email{loock@uni-mainz.de}
\affiliation{Johannes-Gutenberg University of Mainz, Institute of Physics, Staudingerweg 7, 55128 Mainz}

\begin{abstract}
We present two results which combined enable one to reliably detect multimode, multipartite entanglement in the 
presence of measurement errors. The first result leads to a method to compute the best (approximated) physical 
covariance matrix given a measured non-physical one. The other result states that a widely used entanglement condition 
is a consequence of negativity of partial transposition. Our approach can quickly verify entanglement of experimentally 
obtained multipartite states, which is demonstrated on several realistic examples. Compared to existing detection 
schemes, ours is very simple and efficient. In particular, it does not require any complicated optimizations.
\end{abstract}



\maketitle

\textit{Introduction}. Measurement errors are inevitable in real experiments. Not only do they introduce imprecision 
into the measured data, but the key property of physicality of the object under study may even be violated. This is 
well known in quantum state tomography, where the reconstructed density matrix may not be positive semidefinite. 
Techniques like those proposed in Refs.~\cite{nphys.5.27, SciReps.3.3496} are used to clean up the measured data and 
produce physical results. Another instance of this nonphysicality problem and one of the main subjects of the present 
work is related to the covariance matrix (matrix of second-order moments) of a multiparticle or multimode quantum state 
such as an optical, continuous-variable state: The measured covariance matrix may not satisfy the physicality 
condition. 

There are several approaches to tackle this problem in general. One is to model the experimental setup and the 
measurement process to deduce the most probable physical set of data obtained in this process. This is the approach 
taken in Ref.~\cite{nphys.5.27}. However, sometimes we do not have the luxury of knowing the measurement process or its 
model would be exceedingly complicated. In this case we have to recover a physical approximation only from the measured 
data itself without any reference to the process in which this data was obtained.

The natural question to ask is how good is the obtained physical approximation. To give a reliable answer to this 
question we need another piece of information in the form of the strength of the measurement errors. If we know the 
standard deviation, $\sigma$, of the measured quantity from its average value, we can say that our physical 
approximation is good if it fits into a small $\sigma$ interval centered at the average. It remains to be understood how 
small this interval should be.

The acceptable size of this interval depends on the standards adopted by the community in different disciplines. For 
example, in clinical trials a relatively weak $2\sigma$ criterion is used (Refs.~\cite{enc-meas-stat, sacs, 
phys-pharm}, though the full story of clinical studies is more complicated than expressed by the $2\sigma$ criterion). 
On the other hand, in particle physics an informal standard refers to results with the significance $3\sigma$ as 
"evidence" and as a true discovery if the significance is $5\sigma$ \cite{standards, particle}. A recent result 
of fundamental importance is the discovery of gravitational waves, where the significance level was also reported to be 
$5\sigma$ \cite{PhysRevLett.116.061102}. These numbers should not be taken too literally, i.e., that they must be 
exactly three and exactly five. For example, in Tevatron experiments the observation of the top quark was first reported 
as "evidence" with the significance $2.8\sigma$ \cite{PhysRevD.50.2966} and later as a real discovery with the 
significance levels $4.6\sigma$--$4.8\sigma$ \cite{PhysRevLett.74.2632, PhysRevLett.74.2626}. In general, a 
significance level of $s \sigma$ means that to make the wrong conclusion by chance an event outside of an $s \sigma$ 
interval must be realized in the experiment and the probability of such an event quickly decreases as $s$ increases. In 
some cases relatively small values of $s$ are acceptable, but for more fundamental results a stronger confidence is 
needed. We refer to the above results to illustrate that a significance of $5\sigma$ is considered to be sufficient even 
for the discovery of fundamental properties of Nature, and therefore here we shall not impose stronger criteria for 
accepting the experimental results. 

The results that we present here are twofold. First, we propose an algorithm to obtain the best physical approximation 
to a measured non-physical covariance matrix. If, in addition, the standard deviations of the individual matrix elements 
are known then one can also estimate how good this approximation is. The absence of a good physical approximation to 
the measured matrix is then a signature of inaccuracy of the experiment. Our algorithm is based on semidefinite 
optimization and there is a very efficient free software for this (and also a much more general) kind of optimization. 
A low-end desktop PC is enough to perform this algorithm for states with tens of modes.

Another result of our work is to demonstrate that the well-known entanglement condition obtained in 
Ref.~\cite{PhysRevA.67.052315} is, in fact, based on negativity of partial transposition (PT). Combined together, 
the two results give a way to quickly test a multimode quantum state for entanglement even in the presence of 
measurement errors. This approach does not require any optimization and it works surprisingly well; sometimes even 
better than an alternative approach based on genetic optimization proposed in the literature 
\cite{PhysRevLett.114.050501}. The simplicity of our method is very attractive in the high-partite case, where 
otherwise the optimization can take a long time, while taking eigenvectors of matrices, like in our scheme, even with 
thousands of rows and columns is fast on a common PC. 

\begin{figure}
\includegraphics[scale=0.95]{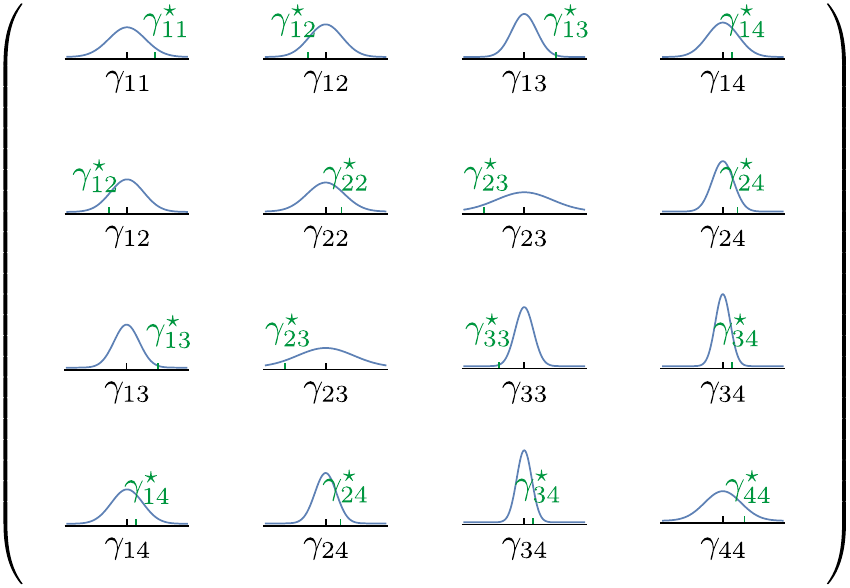}
\caption{(Color online) Covariance matrix of a bipartite state, two-mode state. The measured elements are shown in 
black, the most probable (yet unknown) matrix elements are in green. The shape of the probability distribution of each 
$\gamma_{ij}$ is characterized by the corresponding standard deviation $\sigma_{ij}$, also known from the 
experiment.}\label{fig:1}
\end{figure}

\textit{Measuring covariance matrices}. It is well known that a $2n \times 2n$ real symmetric matrix $\gamma = 
\left(\begin{smallmatrix}\gamma_{xx} & \gamma_{xp} \\ \gamma^{\mathrm{T}}_{xp} & \gamma_{pp} \end{smallmatrix}\right)$ 
is a covariance matrix of an $n$-partite quantum state iff it satisfies either of the two equivalent conditions
\begin{equation}\label{eq:pg}
    \gamma \pm \frac{i}{2} J \geqslant 0,
\end{equation}
where $J = \left(\begin{smallmatrix} 0 & E \\ -E & 0 \end{smallmatrix}\right)$ and $E$ is the $n \times n$ identity 
matrix. The covariance matrices obtained in experiments often violate these conditions, i.e., the matrices on the 
left-hand side of Eq.~\eqref{eq:pg} have small negative eigenvalues. The natural question to ask is: What is the most 
probable physical covariance matrix corresponding to the measured slightly non-physical one? Provided that the 
experiment was performed correctly, the individual measured matrix elements $\gamma_{ij}$ should not differ 
significantly from the matrix elements $\gamma^\star_{ij}$ of the "true" covariance matrix $\gamma^\star$. Thus, we can 
say that the most probable physical matrix is the matrix $\gamma^\star$ with $\max_{i,j} 
|\gamma_{ij}-\gamma^\star_{ij}|$ as small as possible. This is illustrated by Fig.~\ref{fig:1}. If we have standard 
deviations $\sigma_{ij}$ of measurements of the individual matrix elements $\gamma_{ij}$ then the right quantity to 
minimize is $\max_{1 \leqslant i, j \leqslant 2n} |\gamma_{ij}-\gamma^\star_{ij}|/\sigma_{ij}$, so that each true matrix 
element $\gamma^\star_{ij}$ lies in an as small sigma interval around the mean value $\gamma_{ij}$ as possible. The 
solution of this problem can also be used as a test for correctness of the experiment --- when the minimized quantity is 
too large, i.e., there is no physical covariance matrix close to the measured one, then the experiment is likely to have 
been performed inaccurately.

We can now formulate the optimization problem for the best physical approximation of a non-physical covariance matrix 
$\gamma^\circ$:
\begin{equation}\label{eq:gmin}
    \min_{\gamma} \max_{1 \leqslant i \leqslant j \leqslant 2n} 
\frac{|\gamma_{ij}-\gamma^\circ_{ij}|}{\sigma_{ij}},
\end{equation}
where the minimization is over all physical covariance matrices $\gamma$. An optimal solution of this problem 
$\gamma^\star$ is the most probable physical covariance matrix corresponding to the measured matrix $\gamma^\circ$. If 
$\gamma^\circ$ happens to be physical from the very beginning then $\gamma^\star = \gamma^\circ$, as it must be. It 
does not mean that in this case $\gamma^\circ$ is the true physical matrix of the state, it just means it the best 
matrix we obtain from this experiment. 

Note that if we do not have the $\sigma$ matrix, then we can minimize $\max_{i,j}|\gamma_{ij} - \gamma^\circ_{ij}|$, 
which formally coincides with the problem \eqref{eq:gmin} where all $\sigma_{ij}$ are set to 1. We denote the solution 
of this problem by $\tilde{\gamma}^\star$. This means that the same approach works both when we have $\sigma$ and when 
we do not. But in the latter case we will not be able to estimate how good the solution $\tilde{\gamma}^\star$ is.

The problem \eqref{eq:gmin} can be solved by formulating it as a semidefinite optimization problem. The conditions 
\eqref{eq:pg} can be equivalently written in real form as follows:
\begin{equation}\label{eq:pg2}
    \begin{pmatrix}
        \gamma_{xx} & 0 & \gamma_{xp} & \mp \frac{1}{2} E \\[1mm]
        0 & \gamma_{xx} & \pm \frac{1}{2}  E & \gamma_{xp} \\[1mm]
        \gamma^{\mathrm{T}}_{xp} & \pm \frac{1}{2} E & \gamma_{pp} & 0 \\[1mm]
        \mp \frac{1}{2} E & \gamma^{\mathrm{T}}_{xp} & 0 & \gamma_{pp}
    \end{pmatrix}
    \geqslant 0.
\end{equation}
We can also introduce a new real variable, say $s$, which corresponds to the maximum in \eqref{eq:gmin}. Then from the 
condition $\max_{1 \leqslant i \leqslant j \leqslant 2n} |\gamma_{ij} - \gamma^\circ_{ij}|/\sigma_{ij} = s$ it follows 
that the variables of the optimization problem ($s$ and $\gamma_{ij}$) satisfy the following restrictions:
\begin{equation}\label{eq:sg}
    s \sigma_{ij} + \gamma_{ij} \geqslant \gamma^\circ_{ij}, \quad
    s \sigma_{ij} - \gamma_{ij} \geqslant -\gamma^\circ_{ij}.
\end{equation}
Now the condition $s \geqslant 0$, Eq.~\eqref{eq:pg2} and all the inequalities \eqref{eq:sg} can be written as a 
nonnegativity of a single large matrix $M(s, \gamma)$ where $s$ and all individual conditions of 
Eq.~\eqref{eq:sg} are put on the diagonal in addition to the block of Eq.~\eqref{eq:pg2}. This matrix is given 
explicitly in the Appendix. The optimization problem we have to solve reads as
\begin{equation}\label{eq:opt}
    \min_{M(s, \gamma) \geqslant 0} s.
\end{equation}
It is a semidefinite optimization problem with $2n^2 + n + 1$ variables ($s$ and $\gamma$) and the matrix condition of 
size $4n^2 + 6n + 1$. For a block-diagonal $\gamma^\circ$, $\gamma^\circ_{xp} = 0$, the number of variables is $n^2+n+1$ 
and the size of the condition is $2n^2+6n+1$. This problem can be easily solved with an appropriate software.

To give some illustrating examples we refer to Ref.~\cite{PhysRevLett.114.050501}, where three measured 
covariance matrices of four-, six- and ten-partite states were analyzed. We have solved the optimization problem for 
those matrices and found that the minimal values of $s$ for them are 1.88, 0.17 and 0.37, respectively. The 
computation times to solve these problems are: nearly instantly for the four-partite case, around 3 sec for the 
six-partite case and 45 sec for the ten-partite case (on one core of a low-end desktop PC). In the latter case the 
optimization problem has 111 variables and matrix condition of size 261. In the four-partite case the minimal $s$ is 
large enough and nine of the twenty elements of the optimal matrix are $\approx 1.88\sigma$ away from the corresponding 
measured values. One could argue that it might be possible to trade quality of approximations of individual elements 
for the number of elements with large deviations. For example, it might be possible to find a physical matrix where 
only one or two elements lie within larger interval, say $3\sigma$, but the rest are much closer to the measured values. 
In other words, it is the total probability that must be minimized, like the product of probabilities of individual 
elements, not the individual probabilities themselves. Unfortunately, this argument does not work since for a 
correctly performed experiment all the elements must be measured accurately; it does not help if only one or two are 
imprecise but the others are measured perfectly. This is the reason to believe that the experiment with four modes was 
performed far less accurately than the other two, where the results are quite satisfactory. The results are discussed in 
more detail in the Appendix.

Let us compare our approach to finding a physical covariance matrix given a measured nonphysical one to the 
approach used in Ref.~\cite{PhysRevLett.114.050501}. The matrix $\gamma'$ considered there as a physical approximation 
is given by
\begin{equation}\label{eq:gamma'}
    \gamma' = \gamma + 1.001|\lambda_{\mathrm{min}}| E,
\end{equation}
where $\lambda_{\mathrm{min}}$ is the most negative eigenvalue of $\gamma+(i/2)J$. As we noted before, we should look 
for the true physical matrix within small distance of the measured one, so each difference $|\gamma_{ij} - 
\gamma'_{ij}|$ must be a small multiple of $\sigma_{ij}$. The procedure expressed by Eq.~\eqref{eq:gamma'} changes only 
the diagonal elements of the matrix $\gamma$, so computing the ratio $|\gamma_{ii} - \gamma'_{ii}|/\sigma_{ii}$ in the 
four-partite case for the $\gamma_{xx}$ part, we get the values $16.59,\ 6.59,\ 6.29,\ 9.87$. For the $\gamma_{pp}$ 
part, we get $11.84,\ 5.30,\  3.70,\ 11.91$. From the discussion of probabilities in Ref.~\cite{PhysRevA.92.042328} (and 
virtually any textbook on statistics) it follows that the probability of the matrix elements being outside of $11\sigma$ 
interval is smaller than $10^{-27}$, so it is absolutely unrealistic that the matrix $\gamma'$ given by 
Eq.~\eqref{eq:gamma'} was the true covariance matrix in the experiment. Our approach gives a physical covariance matrix 
with much closer elements, but it is still not good enough so that we believe that the four-partite experiment 
contains inaccurate data. For the six-partite case the matrix $\gamma'$ gives a better approximation with only one 
element differing from the corresponding element of $\gamma$ by a quantity larger than $2\sigma$. In the ten-partite 
case some elements differ by $3\sigma$--$4\sigma$ and thus are unlikely to be the true values of the corresponding 
matrix elements. In general, there is no obvious reason to be sure that the expression on the right-hand side of 
Eq.~\eqref{eq:gamma'} produces a good approximation and we have just seen that indeed it may not. A much more reliable 
approach is expressed by the solution of the minimization problem \eqref{eq:gmin}.

\textit{Detecting entanglement}. In Ref.~\cite{PhysRevLett.114.050501} the matrices were also analyzed for 
entangelement by a complicated optimization algorithm. We now demonstrate that comparable (and sometimes even 
better) results can be obtained immediately without any optimization at all. For this purpose, first, we demonstrate 
that a condition of Ref.~\cite{PhysRevA.92.042328} is equivalent to the positivity of partial transposition test of the 
covariance matrix, which is a useful result by itself. For arbitrary real $n$-vectors $\vec{h}$, $\vec{h}'$, $\vec{g}$ 
and $\vec{g}'$ let us introduce the operators $\hat{u} = (\vec{h}, \vec{x})$, $\hat{u}' = (\vec{h}', \vec{x})$, $\hat{v} 
= (\vec{g}, \vec{p})$ and $\hat{v}' = (\vec{g}', \vec{p})$. Then the inequality \eqref{eq:pg2} can be equivalently 
written as follows:
\begin{equation}\label{eq:uvuv}
    \langle (\Delta \hat{u} + \Delta \hat{v}')^2 + (\Delta \hat{u}' + \Delta \hat{v})^2 \rangle 
    \geqslant |(\vec{h}, \vec{g}) - (\vec{h}', \vec{g}')|.
\end{equation}
This condition for arbitrary vectors $\vec{h}$, $\vec{h}'$, $\vec{g}$ and $\vec{g}'$ is equivalent to Eq.~\eqref{eq:pg} 
and thus is necessary and sufficient for physicality of the covariance matrix. Setting $\vec{h}' = \vec{g}' = 0$ we get 
a simpler but weaker condition
\begin{equation}\label{eq:uv}
    \langle (\Delta \hat{u})^2 + (\Delta \hat{v})^2 \rangle \geqslant |(\vec{h}, \vec{g})|.
\end{equation}
This condition is equivalent to the following ones:
\begin{equation}\label{eq:pg3}
    \begin{pmatrix}
        \gamma_{xx} & \pm \frac{1}{2} E \\[1mm]
        \pm \frac{1}{2} E & \gamma_{pp}
    \end{pmatrix}
    \geqslant 0,
\end{equation}
where the matrix on the left-hand side is the central submatrix of the matrix in Eq.~\eqref{eq:pg2}. 

One can get an entanglement condition based on negativity of PT by using a partially transposed matrix 
$\gamma^{\mathrm{PT}}$ in the conditions Eq.~\eqref{eq:uv} or Eq.~\eqref{eq:pg3}. Let us consider a bipartition 
$\mathcal{I} = \{I, J\}$ and partial transposition corresponding to this bipartition. The covariance matrix of the 
partially transposed state can be obtained from the original one by changing the sign of its rows and columns with 
indices that are transposed (the diagonal elements change their sign twice and thus remain positive). Applying the 
condition \eqref{eq:pg3} to this covariance matrix we get the following inequality:
\begin{equation}\label{eq:gpt}
    \begin{pmatrix}
        \gamma_{xx} & \pm\frac{1}{2}E_{\mathcal{I}} \\[1mm]
        \pm\frac{1}{2}E_{\mathcal{I}} & \gamma_{pp}
    \end{pmatrix}
    \geqslant 0,
\end{equation}
which was introduced in Ref.~\cite{PhysRevA.92.042328}. The matrix $E_{\mathcal{I}}$ has $\pm 1$ on the main diagonal 
and the other elements are zero. The diagonal elements with indices from the same group, $I$ or $J$, have the same 
sign. We stress that the two conditions \eqref{eq:pg3}, which differ only by the sign of the off-diagonal block, are 
equivalent and thus only one of them needs to be tested in practice. In vector form the inequalities \eqref{eq:gpt} 
read as $\langle (\Delta \hat{u})^2 + (\Delta \hat{v})^2 \rangle \geqslant |(\vec{h}_I, \vec{g}_I) - (\vec{h}_J, 
\vec{g}_J)|$, where $\vec{h}_I = \{h_{i_1}, \ldots h_{i_k}\}$, $I = \{i_1, \ldots, i_k\}$ and similar notation is used 
for $\vec{h}_J$, $\vec{g}_I$ and $\vec{g}_J$. Combining it with Eq.~\eqref{eq:uv} and noting that $(\vec{h}, \vec{g}) = 
(\vec{h}_I, \vec{g}_I) + (\vec{h}_J, \vec{g}_J)$ we get
\begin{equation}\label{eq:hgpt}
    \langle (\Delta \hat{u})^2 + (\Delta \hat{v})^2 \rangle \geqslant |(\vec{h}_I, \vec{g}_I)| + |(\vec{h}_J, 
\vec{g}_J)|,
\end{equation}
which is exactly the inequality obtained in Ref.~\cite{PhysRevA.67.052315}. This method can be generalized to 
produce similar inequalities for partitions of modes into several groups, not only two. We see that the results of that 
work are just a consequence of the positivity of PT applied to the inequality \eqref{eq:pg3}, which is a 
"part" of the more general inequality \eqref{eq:pg2}. Thus, a more general condition could be obtained from the 
inequality \eqref{eq:uvuv}.

It is not enough just to test partial transpositions of the recovered physical matrix for positive semidefinitness. In 
the presence of the measurement errors we need to verify a stronger statement. Not only should we just violate the 
condition \eqref{eq:hgpt}, where the left-hand side is computed with the recovered physical covariance matrix, but we 
should also violate it with some margin. In other words, we must have
\begin{equation}\label{eq:uvs}
    \frac{|(\vec{h}_I, \vec{g}_I)| + |(\vec{h}_J, \vec{g}_J)| - \langle (\Delta \hat{u})^2 + 
    (\Delta \hat{v})^2 \rangle}{\sigma(\vec{h}, \vec{g})} \geqslant s_0,
\end{equation}
where the denominator of the left-hand side reads as
\begin{equation}
    \sigma(\vec{h}, \vec{g}) = \Bigl(\sum^n_{i,j=1}\sigma^2_{xx, ij} h^2_i h^2_j + 
    \sigma^2_{pp, ij} g^2_i g^2_j\Bigr)^{1/2},
\end{equation}
the standard deviation of the random variable (a linear combination of the random variables $\gamma_{xx, ij}$ and 
$\gamma_{pp, ij}$)
\begin{equation}
    \langle (\Delta \hat{u})^2 + (\Delta \hat{v})^2 \rangle = 
    \sum^n_{i,j=1}\gamma_{xx, ij} h_i h_j + \gamma_{pp, ij} g_i g_j,
\end{equation}
and $s_0$ is the chosen level of confidence, say $s_0 = 3$. If we can find $\vec{h}$ and $\vec{g}$ with such $s_0$, 
than we can be sure that the condition \eqref{eq:hgpt} is really violated. It has been noted in 
Ref.~\cite{PhysRevA.92.042328} that $|(\vec{h}_I, \vec{g}_I)| + |(\vec{h}_J, \vec{g}_J)| = 
\tr\sqrt{\sqrt{X_I}P_I\sqrt{X}_I} + \tr\sqrt{\sqrt{X_J}P_J\sqrt{X}_J}$, where $X = \vec{h}\vec{h}^{\mathrm{T}}$, $P = 
\vec{g}\vec{g}^{\mathrm{T}}$ are rank-1 matrices and $X_I$ is the submatrix of $X$ whose row and column indicies are in 
$I$. This means that the quantity \eqref{eq:uvs} is a special case of the one used in 
Ref.~\cite{PhysRevLett.114.050501} to detect entanglement. 

The proof of the inequality \eqref{eq:hgpt} suggests a simple way to choose vectors $\vec{h}$ and $\vec{g}$ to have 
large $s_0$ in Eq.~\eqref{eq:uvs}. Suppose that the matrices \eqref{eq:gpt} have negative eigenvalues (which means that 
partial transposition of the state is negative) and $\vec{z} = (z_1, \ldots, z_{2n})$ is the eigenvector with the most 
negative eigenvalue. We can try to use vectors $(z_1, \ldots, z_n)$ and $(z_{n+1}, \ldots, z_{2n})$ as the vectors 
$\vec{h}$ and $\vec{g}$. There is no guarantee that these vectors will always satisfy the inequality \eqref{eq:uvs} with 
large $s_0$, but it is a good starting point. If they satisfy that inequality, we are ready, if they do not, we will 
have to perform a more complicated analysis.

To illustrate the applicability of this approach, let us apply it to the three experimentally measured covariance 
matrices considered before. We consider only bipartitions, since if the bipartitions pass the test then there is no 
reason to test other partitions. We have also applied our method to the matrix $\gamma'$, Eq.~\eqref{eq:gamma'}, which 
we know could not be the true matrix in the experiment, but we included it to demonstrate that our approach gives 
comparable results without performing any optimization. Just to compare the results we give the level of violation for 
the matrix $\tilde{\gamma}^\star$ obtained as a solution of the optimization problem \eqref{eq:gmin} where the errors 
$\sigma$ are not available. All these results are given in the Appendix. We see that our approach applied to the matrix 
$\gamma'$ produces comparable violations which are just marginally smaller than those obtained with optimization, and 
this difference is completely irrelevant for such large violations. For the ten-partite case the minimal violation was 
reported to be 1.1 for the bipartition $1,10|23456789$. If we apply our method to the matrix $\gamma'$ we get the 
violation 2.6. In this case our approach immediately produces a much better result than that obtained with the 
optimization approach of Ref.~\cite{PhysRevLett.114.050501}. For the matrix $\gamma^\star$, which is the most probably 
true covariance matrix in the experiment, this violation is 3.6. Using the technique of Ref.~\cite{PhysRevA.92.042328} 
we are even able to verify that the four-partite state is genuine multipartite entangled with confidence level 3 (see 
Appendix).

\textit{Conclusion}. In conclusion, we have presented an algorithm to recover the best physical approximation to the 
experimentally measured covariance matrix and applied it to some realistic measured data. The algorithm is based on 
semidefinite optimization and can be easily implemented by using free optimization software. In addition, we have proved 
that a widely used entanglement condition is in fact based on the negativity of PT and can be easily checked by testing 
the recovered covariance matrix of the partially transposed state for negative eigenvalues. It has been shown that the 
eigenvector corresponding to the most negative eigenvalue can be a good entanglement witness even in the presence of 
measurement errors and sometimes it is even better than one obtained with complicated optimization algorithms (like in 
the ten-partite case considered in Ref.~\cite{PhysRevLett.114.050501}). Applying the technique of our previous work to 
the recovered covariance matrix of the four-partite state under study we demonstrate that that state is genuine 
multipartite entangled.

\appendix

\section{Appendix: Semidefinite optimization problem}

Here we give all the details missing in the main part of the work. First, we provide a detailed discussion of the 
optimization problem \eqref{eq:opt}. The condition $M(s, \gamma) \geqslant 0$ is equivalent to the condition 
$M_1(s,\gamma) \leqslant M_0$, where the matrices $M_0$ and $M_1(s,\gamma)$ in the singlemode case read as
\begin{widetext}
\setcounter{MaxMatrixCols}{11}
\begin{equation}
    M_1(s,\gamma) = -
    \begin{pmatrix}
        s & 0 & 0 & 0 & 0 & 0 \\[1mm]
        0 & \gamma_{xx} & 0 & \gamma_{xp} & 0 & 0 & 0 & 0 & 0 & 0 & 0 \\[1mm]
        0 & 0 & \gamma_{xx} & 0 & \gamma_{xp} & 0 & 0 & 0 & 0 & 0 & 0 \\[1mm]
        0 & \gamma_{xp} & 0 & \gamma_{pp} & 0 & 0 & 0 & 0 & 0 & 0 & 0 \\[1mm]
        0 & 0 & \gamma_{xp} & 0 & \gamma_{pp} & 0 & 0 & 0 & 0 & 0 & 0 \\[1mm]
        0 & 0 & 0 & 0 & 0 & s \sigma_{xx} + \gamma_{xx} & 0 & 0 & 0 & 0 & 0 \\[1mm]
        0 & 0 & 0 & 0 & 0 & 0 & s \sigma_{xx} - \gamma_{xx} & 0 & 0 & 0 & 0 \\[1mm]
        0 & 0 & 0 & 0 & 0 & 0 & 0 & s \sigma_{xp} + \gamma_{xp} & 0 & 0 & 0 \\[1mm]
        0 & 0 & 0 & 0 & 0 & 0 & 0 & 0 & s \sigma_{xp} - \gamma_{xp} & 0 & 0 \\[1mm]
        0 & 0 & 0 & 0 & 0 & 0 & 0 & 0 & 0 & s \sigma_{pp} + \gamma_{pp} & 0 \\[1mm]
        0 & 0 & 0 & 0 & 0 & 0 & 0 & 0 & 0 & 0 & s \sigma_{pp} - \gamma_{pp}
    \end{pmatrix},
\end{equation}
\begin{equation}
    M_0 = 
    \begin{pmatrix}
        0 & 0 & 0 & 0 & 0 & 0 & 0 & 0 & 0 & 0 & 0 \\[1mm]
        0 & 0 & 0 & 0 & -\frac{1}{2} & 0 & 0 & 0 & 0 & 0 & 0 \\[1mm]
        0 & 0 & 0 & \frac{1}{2} & 0 & 0 & 0 & 0 & 0 & 0 & 0 \\[1mm]
        0 & 0 & \frac{1}{2} & 0 & 0 & 0 & 0 & 0 & 0 & 0 & 0 \\[1mm]
        0 & -\frac{1}{2} & 0 & 0 & 0 & 0 & 0 & 0 & 0 & 0 & 0 \\[1mm]
        0 & 0 & 0 & 0 & 0 & -\gamma^\circ_{xx} & 0 & 0 & 0 & 0 & 0 \\[1mm]
        0 & 0 & 0 & 0 & 0 & 0 & \gamma^\circ_{xx} & 0 & 0 & 0 & 0 \\[1mm]
        0 & 0 & 0 & 0 & 0 & 0 & 0 & -\gamma^\circ_{xp} & 0 & 0 & 0 \\[1mm]
        0 & 0 & 0 & 0 & 0 & 0 & 0 & 0 & \gamma^\circ_{xp} & 0 & 0 \\[1mm]
        0 & 0 & 0 & 0 & 0 & 0 & 0 & 0 & 0 & -\gamma^\circ_{pp} & 0 \\[1mm]
        0 & 0 & 0 & 0 & 0 & 0 & 0 & 0 & 0 & 0 & \gamma^\circ_{pp}
    \end{pmatrix}.
\end{equation}
\end{widetext}
The structure of these matrices in the multimode case is the same, but the size is much larger. We stress that the 
parameters $\gamma^\circ$ and $\sigma$ in these matrices are constants (they are the data obtained in an experiment) and 
the variables of the optimization problem are $s$ and the elements of $\gamma$. The optimization problem now reads as
\begin{equation}\label{eq:opt4s}
    \max_{M_1(s,\gamma) \leqslant M_0} s,
\end{equation}
and it is a semidefinite optimization problem. The matrices $M_0$ and $M_1$ are sparse, which allows to compactly 
represent them even for states with hundreds of modes. This semidefinite optimization problem can be solved with an 
appropriate software. The software of our choice is \textsl{cvxopt}\footnote{\texttt{http://cvxopt.org}}. 

Now we give the details about the three cases considered in the main part. The measured covariance matrix in the 
four-partite case is
\begin{equation}
\begin{split}
    \gamma^\circ_{xx} &= 
    \begin{pmatrix}
        1.09921 & 0.16092 & -0.17608 & -0.84831 \\
        0.16092 & 0.40938 & -0.1606 & -0.18963 \\
        -0.17608 & -0.16060 & 0.46060 & 0.04318 \\
        -0.84831 & -0.18963 & 0.04318 & 1.064185
    \end{pmatrix} \\
    \gamma^\circ_{pp} &=
    \begin{pmatrix}
        1.09921 & 0.35533 & 0.36439 & 0.91384 \\
        0.35533 & 0.92282 & 0.57439 & 0.43388 \\
        0.36439 & 0.57439 & 1.04339 & 0.34868 \\
        0.91384 & 0.43388 & 0.34868 & 1.06419
    \end{pmatrix},
\end{split}
\end{equation}
and $\gamma^\circ_{xp} = 0$. The standard deviations are given by the matrices
\begin{equation}
\begin{split}
    \sigma_{xx} &= 
    \begin{pmatrix}
        0.00326 & 0.01041 & 0.00893 & 0.00646 \\
        0.01041 & 0.00822 & 0.01847 & 0.01899 \\
        0.00893 & 0.01847 & 0.00861 & 0.01345 \\
        0.00646 & 0.01899 & 0.01345 & 0.00549
    \end{pmatrix} \\
    \sigma_{pp} &= 
    \begin{pmatrix}
        0.00457 & 0.01009 & 0.02767 & 0.04288 \\
        0.01009 & 0.01022 & 0.02100 & 0.02085 \\
        0.02767 & 0.02100 & 0.01465 & 0.01955 \\
        0.04288 & 0.02085 & 0.01955 & 0.00455
    \end{pmatrix}.
\end{split}
\end{equation}
The solution of the problem \eqref{eq:opt4s} for this data is
\begin{equation}\label{eq:opt4}
\begin{split}
    \gamma^{\star}_{xx} &=
    \begin{pmatrix}
        1.10535 & 0.14133 & -0.16983 & -0.84598 \\
        0.14133 & 0.42485 & -0.19067 & -0.19569 \\
        -0.16983 & -0.19067 & 0.46261 & 0.04967 \\
        -0.84598 & -0.19569 & 0.04967 & 1.06482
    \end{pmatrix} \\
    \gamma^{\star}_{pp} &=
    \begin{pmatrix}
        1.10782 & 0.33634 & 0.41646 & 0.89981 \\
        0.33634 & 0.94206 & 0.53487 & 0.40398 \\
        0.41646 & 0.53487 & 1.07097 & 0.36679 \\
        0.89981 & 0.40398 & 0.36679 & 1.06472
    \end{pmatrix}.
\end{split}
\end{equation}
We have computed the eigenvector corresponding to the most negative eigenvalue for three different matrices --- 
$\gamma'$ used in Ref.~\cite{PhysRevLett.114.050501}, the optimal solution $\gamma^\star$ given by Eq.~\eqref{eq:opt4}, 
and the optimal solution $\tilde{\gamma}^\star$ obtained without taking $\sigma$ into account --- and checked the 
significance $s_0$ these vectors provide in Eq.~\eqref{eq:uvs}. The results are shown in Table~\ref{tbl:4}. It can be 
seen that these vectors give results comparable to those obtained with sophisticated optimization, and the difference 
is absolutely irrelevant for such high $s_0$.

\begin{table}[ht]
\begin{tabular}{|c|D{.}{.}{3.3}|D{.}{.}{3.3}|D{.}{.}{3.3}|D{.}{.}{3.3}|}
\toprule[0.6pt]
 \multirow{2}{*}{Bipartition} & \multicolumn{4}{c|}{Violation} \\ 
\cline{2-5}
 & 
\multicolumn{1}{c|}{\textrm{Ref.~\cite{PhysRevLett.114.050501}}} & 
\multicolumn{1}{c|}{$\gamma'$} & 
\multicolumn{1}{c|}{$\gamma^\star$} & 
\multicolumn{1}{c|}{$\tilde{\gamma}^\star$} \\
\midrule[0.4pt]
$1|234$ & 20.93 & 16.18 & 19.09 & 18.46 \\ 
$2|134$ & 13.17 & 13.04 & 16.52 & 15.93 \\ 
$3|124$ & 11.21 & 11.18 & 15.42 & 14.52 \\ 
$4|123$ & 21.06 & 16.24 & 19.27 & 18.96 \\ 
$12|34$ & 24.34 & 18.14 & 21.69 & 20.77 \\ 
$13|24$ & 23.52 & 15.97 & 18.87 & 18.44 \\ 
$14|23$ & 4.66  & 4.29  & 8.48  & 7.57 \\
\bottomrule[0.6pt]
\end{tabular}
\caption{Comparison of the confidence level $s_0$, Eq.~\eqref{eq:uvs}, for the four-partite case.}\label{tbl:4}
\end{table}

A similar comparison for the six-partite case is given in Table~\ref{tbl:6}. Here we see the same behavior --- our 
eigenvector approach gives slightly smaller but comparable violations. The violations are even larger than in the 
previous four-partite case, and this difference plays no role, especially taken into account that computing eigenvalues 
of such a small matrix costs nothing. For the ten-partite case we cannot give full comparison, because the results are 
not provided in Ref.~\cite{PhysRevLett.114.050501}, but it has been mentioned in the main part that for at least one 
bipartition our approach gives substantially stronger violation than the optimization algorithms implemented in that 
work. We stress that taking eigenvalues is a very cheap operation and can be efficiently performed for matrices with 
thousands rows and columns even on a low-end desktop PC. So, our approach is a very simple yet effective method that 
can immediately detect entangelement of states with huge number of parts even in the case of imperfect measurements. In 
rare cases where this approach does not work one can utilize a more expensive optimization technique. 

\begin{table}[ht]
\begin{tabular}{|c|D{.}{.}{3.3}|D{.}{.}{3.3}|D{.}{.}{3.3}|D{.}{.}{3.3}|}
\toprule[0.6pt]
 \multirow{2}{*}{Bipartition} & \multicolumn{4}{c|}{Violation} \\ \cline{2-5}
 & \multicolumn{1}{c|}{\textrm{Ref.~\cite{PhysRevLett.114.050501}}} & 
\multicolumn{1}{c|}{$\gamma'$} & \multicolumn{1}{c|}{$\gamma^\star$} & 
\multicolumn{1}{c|}{$\tilde{\gamma}^\star$} \\
\midrule[0.4pt]
$1|23456$ & 40.09 & 38.47 & 39.05 & 38.81 \\[1mm]
$2|13456$ & 36.18 & 35.48 & 36.45 & 36.23 \\[1mm]
$3|12456$ & 20.27 & 19.32 & 19.93 & 19.81 \\[1mm]
$4|12356$ & 20.01 & 18.69 & 19.14 & 19.06 \\[1mm]
$5|12346$ & 27.15 & 26.66 & 27.42 & 27.28 \\[1mm]
$6|12345$ & 49.22 & 45.70 & 46.34 & 46.09 \\[1mm]
$12|3456$ & 53.54 & 48.79 & 49.50 & 49.23 \\[1mm]
$13|2456$ & 45.57 & 42.28 & 42.81 & 42.56 \\[1mm]
$14|2356$ & 44.79 & 39.85 & 40.44 & 40.24 \\[1mm]
$15|2346$ & 45.28 & 38.01 & 38.51 & 38.34 \\[1mm]
$16|2345$ & 31.18 & 29.99 & 30.87 & 30.68 \\[1mm]
$23|1456$ & 40.16 & 38.78 & 39.68 & 39.44 \\[1mm]
$24|1356$ & 37.70 & 35.89 & 36.84 & 36.60 \\[1mm]
$25|1346$ & 35.26 & 31.18 & 32.07 & 31.85 \\[1mm]
$26|1345$ & 47.02 & 40.27 & 40.80 & 40.63 \\[1mm]
$34|1256$ & 24.83 & 21.49 & 22.14 & 22.01 \\[1mm]
$35|1246$ & 28.79 & 25.05 & 25.74 & 25.69 \\[1mm]
$36|1245$ & 50.19 & 45.09 & 45.73 & 45.51 \\[1mm]
$45|1236$ & 30.63 & 28.26 & 28.93 & 28.81 \\[1mm]
$46|1235$ & 51.50 & 47.23 & 47.82 & 47.58 \\[1mm]
$56|1234$ & 56.08 & 51.54 & 52.24 & 51.95 \\[1mm]
$123|456$ & 56.66 & 52.09 & 52.76 & 52.49 \\[1mm]
$124|356$ & 54.40 & 49.40 & 50.12 & 49.86 \\[1mm]
$125|346$ & 50.65 & 45.55 & 46.17 & 45.95 \\[1mm]
$126|345$ & 28.96 & 25.71 & 26.47 & 26.39 \\[1mm]
$134|256$ & 47.68 & 42.95 & 43.53 & 43.29 \\[1mm]
$135|246$ & 47.28 & 40.51 & 40.99 & 40.82 \\[1mm]
$136|245$ & 34.24 & 29.69 & 30.60 & 30.38 \\[1mm]
$145|236$ & 47.33 & 38.89 & 39.40 & 39.27 \\[1mm]
$146|235$ & 35.34 & 33.31 & 34.14 & 33.94 \\[1mm]
$156|234$ & 39.49 & 38.47 & 39.39 & 39.13 \\
\bottomrule[0.6pt]
\end{tabular}
\caption{Comparison of the confidence level $s_0$, Eq.~\eqref{eq:uvs}, for the six-partite case.}\label{tbl:6}
\end{table}

Our technique is not limited to detect only simple kind of entanglement, we are also able to detect genuine 
multipartite entanglement in some cases. This is substantially more complicated problem and our approach works for 
small number of parts. The goal is to violate all the equations \eqref{eq:uvs} simultaneously for all bipartitions. 
Using the method of Ref.~\cite{PhysRevA.92.042328} we have found the following pair of matrices:
\begin{equation}
\begin{split}
    X &= 
    \begin{pmatrix}
        0.29331 & 0.03784 & 0.22823 & 0.23107 \\
        0.03784 & 0.58693 & 0.17803 & 0.17187 \\
        0.22823 & 0.17803 & 0.38153 & 0.19831 \\
        0.23107 & 0.17187 & 0.19831 & 0.28106
    \end{pmatrix} \\
    P &= 
    \begin{pmatrix}
        0.20468 & -0.00241 & -0.08516 & -0.10549 \\
        -0.00241 & 0.36480 & -0.15864 & -0.13005 \\
        -0.08516 & -0.15864 & 0.23421 & 0.01436 \\
        -0.10549 & -0.13005 & 0.01436 & 0.21535    
    \end{pmatrix}
\end{split}
\end{equation}
These matrices are a genuine entanglement witness for the four-partite state under study or, more precisely, for the 
state with the covariance matrix given by Eq.~\eqref{eq:opt4}. In fact, for the bipartition $1|234$ the maximum is 
attained at 
\begin{equation}
\begin{split}
    X' &= 
    \begin{pmatrix}
        0.29331 & \tikzmarkin[hor=style orange]{el1}0.11794 & 0.14763 & 0.11827 \tikzmarkend{el1} \\
        \tikzmarkin[ver=style orange]{el2} 0.11794 & 0.58693 & 0.17803 & 0.17187 \\
        0.14763 & 0.17803 & 0.38153 & 0.19831 \\
        0.11827\tikzmarkend{el2} & 0.17187 & 0.19831 & 0.28106
    \end{pmatrix} \\
    P' &= 
    \begin{pmatrix}
        0.20468 & \tikzmarkin[hor=style orange]{el3}0.01396 & 0.05572 & 0.04133 \tikzmarkend{el3} \\
        \tikzmarkin[ver=style orange]{el4}0.01396 & 0.36480 & -0.15863 & -0.13005 \\
        0.05572 & -0.15864 & 0.23421 & 0.01436 \\
        0.04133\tikzmarkend{el4} & -0.13005 & 0.01436 & 0.21535 
    \end{pmatrix}
    \nonumber
\end{split}
\end{equation}
and the violation is equal to 
\begin{equation}
    \frac{\mathcal{B}_{1|234}(X, P) - \mathcal{G}(X, P)}{\sigma(X, P)} \approx 5.57.
\end{equation}
For the bipartition $2|134$ the maximum is attained at 
\begin{equation}
\begin{split}
    X' &= 
    \begin{pmatrix}
        0.29331 & \tikzmarkin[hor=style orange]{el5}0.12608\tikzmarkend{el5} & 0.22823 & 0.23107 \\
        \tikzmarkin[hor=style orange]{el6}0.12608\tikzmarkend{el6} & 0.58693 & \tikzmarkin[hor=style orange]{el7}0.08149 
& 0.08280\tikzmarkend{el7} \\
        0.22823 & \tikzmarkin[ver=style orange]{el8}0.08149 & 0.38153 & 0.19831 \\
        0.23107 & 0.08280\tikzmarkend{el8} & 0.19831 & 0.28106
    \end{pmatrix} \\
    P' &= 
    \begin{pmatrix}
        0.20468 & \tikzmarkin[hor=style orange]{el9}0.06250\tikzmarkend{el9} & -0.08516 & -0.10549 \\
        \tikzmarkin[hor=style orange]{el10}0.06250\tikzmarkend{el10} & 0.36480 & \tikzmarkin[hor=style 
orange]{el11}0.01133 & 0.00155\tikzmarkend{el11} \\
        -0.08516 & \tikzmarkin[ver=style orange]{el12}0.01133 & 0.23421 & 0.01436 \\
        -0.10549 & 0.00155\tikzmarkend{el12} & 0.01436 & 0.21535 
    \end{pmatrix}
    \nonumber
\end{split}
\end{equation}
and the violation is equal to 
\begin{equation}
    \frac{\mathcal{B}_{2|134}(X, P) - \mathcal{G}(X, P)}{\sigma(X, P)} \approx 3.04.
\end{equation}
For the bipartition $3|124$ the maximum is attained at
\begin{equation}
\begin{split}
    X' &= 
    \begin{pmatrix}
        0.29331 & 0.03784 & \tikzmarkin[ver=style orange]{el13}0.17402 & 0.23107 \\
        0.03784 & 0.58693 & 0.09827\tikzmarkend{el13} & 0.17187 \\
        \tikzmarkin[hor=style orange]{el14}0.17402 & 0.09827\tikzmarkend{el14} & 0.38153 & \tikzmarkin[hor=style 
orange]{el15}0.22962\tikzmarkend{el15} \\
        0.23107 & 0.17187 & \tikzmarkin[hor=style orange]{el16}0.22962\tikzmarkend{el16} & 0.28106 
    \end{pmatrix} \\
    P' &= 
    \begin{pmatrix}
        0.20468 & -0.00241 & \tikzmarkin[ver=style orange]{el17}0.01222 & -0.10549 \\
        -0.00241 & 0.36480 & -0.00151\tikzmarkend{el17} & -0.13005 \\
        \tikzmarkin[hor=style orange]{el18}0.01222 & -0.00151\tikzmarkend{el18} & 0.23421 & \tikzmarkin[hor=style 
orange]{el19}0.11148\tikzmarkend{el19} \\
        -0.10549 & -0.13005 & \tikzmarkin[hor=style orange]{el20}0.11148\tikzmarkend{el20} & 0.21535 
    \end{pmatrix}
    \nonumber
\end{split}
\end{equation}
and the violation is equal to 
\begin{equation}
    \frac{\mathcal{B}_{3|124}(X, P) - \mathcal{G}(X, P)}{\sigma(X, P)} \approx 3.05.
\end{equation}
For the bipartition $4|123$ the maximum is attained at
\begin{equation}
\begin{split}
    X' &= 
    \begin{pmatrix}
        0.29331 & 0.03784 & 0.22823 & \tikzmarkin[ver=style orange]{el22}0.13781 \\
        0.03784 & 0.58693 & 0.17803 & 0.10005 \\
        0.22823 & 0.17803 & 0.38153 & 0.22978\tikzmarkend{el22} \\
        \tikzmarkin[hor=style orange]{el21}0.13781 & 0.10005 & 0.22978\tikzmarkend{el21} & 0.28106  
    \end{pmatrix} \\
    P' &= 
    \begin{pmatrix}
        0.20468 & -0.00241 & -0.08516 & \tikzmarkin[ver=style orange]{el24}0.01806 \\
        -0.00241 & 0.36480 & -0.15864 & -0.00107 \\
        -0.08516 & -0.15864 & 0.23421 & 0.10894\tikzmarkend{el24} \\
        \tikzmarkin[hor=style orange]{el23}0.01806 & -0.00107 & 0.10894\tikzmarkend{el23} & 0.21535 
    \end{pmatrix}
    \nonumber
\end{split}
\end{equation}
and the violation is equal to 
\begin{equation}
    \frac{\mathcal{B}_{4|123}(X, P) - \mathcal{G}(X, P)}{\sigma(X, P)} \approx 5.03.
\end{equation}
For the bipartition $12|34$ the maximum is attained at
\begin{equation}
\begin{split}
    X' &= 
    \begin{pmatrix}
        0.29331 & 0.03784 & \tikzmarkin[ver=style orange]{el25}0.14044 & 0.11243 \\
        0.03784 & 0.58693 & 0.08302 & 0.07887\tikzmarkend{el25} \\
        \tikzmarkin[ver=style orange]{el26}0.14044 & 0.08302 & 0.38153 & 0.19831 \\
        0.11243 & 0.07887\tikzmarkend{el26} & 0.19831 & 0.28106   
    \end{pmatrix} \\
    P' &= 
    \begin{pmatrix}
        0.20468 & -0.00241 & \tikzmarkin[ver=style orange]{el27}0.07716 & 0.06071 \\
        -0.00241 & 0.36480 & 0.03752 & 0.04237\tikzmarkend{el27} \\
        \tikzmarkin[ver=style orange]{el28}0.07716 & 0.03752 & 0.23421 & 0.01436 \\
        0.06071 & 0.04237\tikzmarkend{el28} & 0.01436 & 0.21535  
    \end{pmatrix}
    \nonumber
\end{split}
\end{equation}
and the violation is equal to 
\begin{equation}
    \frac{\mathcal{B}_{12|34}(X, P) - \mathcal{G}(X, P)}{\sigma(X, P)} \approx 15.28.
\end{equation}
For the bipartition $13|24$ the maximum is attained at
\begin{equation}
\begin{split}
    X' &= 
    \begin{pmatrix}
        0.29331 & \tikzmarkin[hor=style orange]{el29}0.10141\tikzmarkend{el29} & 0.22823 & \tikzmarkin[hor=style 
orange]{el30}0.14157\tikzmarkend{el30} \\
        \tikzmarkin[hor=style orange]{el31}0.10141\tikzmarkend{el31} & 0.58693 & \tikzmarkin[hor=style 
orange]{el32}0.09542\tikzmarkend{el32} & 0.17187 \\
        0.22823 & \tikzmarkin[hor=style orange]{el33}0.09542\tikzmarkend{el33} & 0.38153 & \tikzmarkin[hor=style 
orange]{el34}0.21426\tikzmarkend{el34} \\
        \tikzmarkin[hor=style orange]{el35}0.14157\tikzmarkend{el35} & 0.17187 & \tikzmarkin[hor=style 
orange]{el36}0.21426\tikzmarkend{el36} & 0.28106   
    \end{pmatrix} \\
    P' &= 
    \begin{pmatrix}
        0.20468 & \tikzmarkin[hor=style orange]{el37}0.03763\tikzmarkend{el37} & -0.08516 & \tikzmarkin[hor=style 
orange]{el38}0.00885\tikzmarkend{el38} \\
        \tikzmarkin[hor=style orange]{el39}0.03763\tikzmarkend{el39} & 0.36480 & \tikzmarkin[hor=style 
orange]{el40}-0.01959\tikzmarkend{el40} & -0.13005 \\
        -0.08516 & \tikzmarkin[hor=style orange]{el41}-0.01959\tikzmarkend{el41} & 0.23421 & \tikzmarkin[hor=style 
orange]{el42}0.11690\tikzmarkend{el42} \\
        \tikzmarkin[hor=style orange]{el43}0.00885\tikzmarkend{el43} & -0.13005 & \tikzmarkin[hor=style 
orange]{el44}0.11690\tikzmarkend{el44} & 0.21535   
    \end{pmatrix}
    \nonumber
\end{split}
\end{equation}
and the violation is equal to 
\begin{equation}
    \frac{\mathcal{B}_{13|24}(X, P) - \mathcal{G}(X, P)}{\sigma(X, P)} \approx 6.66.
\end{equation}
For the bipartition $14|23$ the maximum is attained at
\begin{equation}
\begin{split}
    X' &= 
    \begin{pmatrix}
        0.29331 & \tikzmarkin[hor=style orange]{el45}0.10225 & 0.17399\tikzmarkend{el45} & 0.23107 \\
        \tikzmarkin[ver=style orange]{el47}0.10225 & 0.58693 & 0.17803 & \tikzmarkin[ver=style orange]{el48}0.09153 \\
        0.17399\tikzmarkend{el47} & 0.17803 & 0.38153 & 0.21520\tikzmarkend{el48} \\
        0.23107 & \tikzmarkin[hor=style orange]{el46}0.09153 & 0.21520\tikzmarkend{el46} & 0.28106    
    \end{pmatrix} \\
    P' &= 
    \begin{pmatrix}
        0.20468 & \tikzmarkin[hor=style orange]{el49}0.03253 & 0.00714\tikzmarkend{el49} & -0.10549 \\
        \tikzmarkin[ver=style orange]{el51}0.03253 & 0.36480 & -0.15864 & \tikzmarkin[ver=style orange]{el52}-0.02570 \\
        0.00714\tikzmarkend{el51} & -0.15864 & 0.23421 & 0.10579\tikzmarkend{el52} \\
        -0.10549 & \tikzmarkin[hor=style orange]{el50}-0.02570 & 0.10579\tikzmarkend{el50} & 0.21535    
    \end{pmatrix}
    \nonumber
\end{split}
\end{equation}
and the violation is equal to 
\begin{equation}
    \frac{\mathcal{B}_{14|23}(X, P) - \mathcal{G}(X, P)}{\sigma(X, P)} \approx 3.04.
\end{equation}
We thus have that for all bipartitions $\mathcal{I}$ the inequality 
\begin{equation}
    \frac{\mathcal{B}_{\mathcal{I}}(X, P) - \mathcal{G}(X, P)}{\sigma(X, P)} > 3
\end{equation}
is satisfied, so we can safely claim that the four-partite state under study is genuine entangled (provided that the 
experiment has been performed correctly).


\begin{thebibliography}{15}%
\makeatletter
\providecommand \@ifxundefined [1]{%
 \@ifx{#1\undefined}
}%
\providecommand \@ifnum [1]{%
 \ifnum #1\expandafter \@firstoftwo
 \else \expandafter \@secondoftwo
 \fi
}%
\providecommand \@ifx [1]{%
 \ifx #1\expandafter \@firstoftwo
 \else \expandafter \@secondoftwo
 \fi
}%
\providecommand \natexlab [1]{#1}%
\providecommand \enquote  [1]{``#1''}%
\providecommand \bibnamefont  [1]{#1}%
\providecommand \bibfnamefont [1]{#1}%
\providecommand \citenamefont [1]{#1}%
\providecommand \href@noop [0]{\@secondoftwo}%
\providecommand \href [0]{\begingroup \@sanitize@url \@href}%
\providecommand \@href[1]{\@@startlink{#1}\@@href}%
\providecommand \@@href[1]{\endgroup#1\@@endlink}%
\providecommand \@sanitize@url [0]{\catcode `\\12\catcode `\$12\catcode
  `\&12\catcode `\#12\catcode `\^12\catcode `\_12\catcode `\%12\relax}%
\providecommand \@@startlink[1]{}%
\providecommand \@@endlink[0]{}%
\providecommand \url  [0]{\begingroup\@sanitize@url \@url }%
\providecommand \@url [1]{\endgroup\@href {#1}{\urlprefix }}%
\providecommand \urlprefix  [0]{URL }%
\providecommand \Eprint [0]{\href }%
\providecommand \doibase [0]{http://dx.doi.org/}%
\providecommand \selectlanguage [0]{\@gobble}%
\providecommand \bibinfo  [0]{\@secondoftwo}%
\providecommand \bibfield  [0]{\@secondoftwo}%
\providecommand \translation [1]{[#1]}%
\providecommand \BibitemOpen [0]{}%
\providecommand \bibitemStop [0]{}%
\providecommand \bibitemNoStop [0]{.\EOS\space}%
\providecommand \EOS [0]{\spacefactor3000\relax}%
\providecommand \BibitemShut  [1]{\csname bibitem#1\endcsname}%
\let\auto@bib@innerbib\@empty
\bibitem [{\citenamefont {Lundeen}\ \emph {et~al.}(2009)\citenamefont
  {Lundeen}, \citenamefont {Feito}, \citenamefont {Coldenstrodt-Ronge},
  \citenamefont {Pregnell}, \citenamefont {Silberhorn}, \citenamefont {Ralph},
  \citenamefont {Eisert}, \citenamefont {Plenio},\ and\ \citenamefont
  {Walmsley}}]{nphys.5.27}%
  \BibitemOpen
  \bibfield  {author} {\bibinfo {author} {\bibfnamefont {J.~S.}\ \bibnamefont
  {Lundeen}}, \bibinfo {author} {\bibfnamefont {A.}~\bibnamefont {Feito}},
  \bibinfo {author} {\bibfnamefont {H.}~\bibnamefont {Coldenstrodt-Ronge}},
  \bibinfo {author} {\bibfnamefont {K.~L.}\ \bibnamefont {Pregnell}}, \bibinfo
  {author} {\bibfnamefont {C.}~\bibnamefont {Silberhorn}}, \bibinfo {author}
  {\bibfnamefont {T.~C.}\ \bibnamefont {Ralph}}, \bibinfo {author}
  {\bibfnamefont {J.}~\bibnamefont {Eisert}}, \bibinfo {author} {\bibfnamefont
  {M.~B.}\ \bibnamefont {Plenio}}, \ and\ \bibinfo {author} {\bibfnamefont
  {I.~A.}\ \bibnamefont {Walmsley}},\ }\href@noop {} {\bibfield  {journal}
  {\bibinfo  {journal} {Nature Physics}\ }\textbf {\bibinfo {volume} {5}},\
  \bibinfo {pages} {27} (\bibinfo {year} {2009})}\BibitemShut {NoStop}%
\bibitem [{\citenamefont {Qi}\ \emph {et~al.}(2013)\citenamefont {Qi},
  \citenamefont {Hou}, \citenamefont {Li}, \citenamefont {Dong}, \citenamefont
  {Xiang},\ and\ \citenamefont {Guo}}]{SciReps.3.3496}%
  \BibitemOpen
  \bibfield  {author} {\bibinfo {author} {\bibfnamefont {B.}~\bibnamefont
  {Qi}}, \bibinfo {author} {\bibfnamefont {Z.}~\bibnamefont {Hou}}, \bibinfo
  {author} {\bibfnamefont {L.}~\bibnamefont {Li}}, \bibinfo {author}
  {\bibfnamefont {D.}~\bibnamefont {Dong}}, \bibinfo {author} {\bibfnamefont
  {G.}~\bibnamefont {Xiang}}, \ and\ \bibinfo {author} {\bibfnamefont
  {G.}~\bibnamefont {Guo}},\ }\href@noop {} {\bibfield  {journal} {\bibinfo
  {journal} {Scientific Reports}\ }\textbf {\bibinfo {volume} {3}},\ \bibinfo
  {pages} {3496} (\bibinfo {year} {2013})}\BibitemShut {NoStop}%
\bibitem [{\citenamefont {Salkind}(2006)}]{enc-meas-stat}%
  \BibitemOpen
  \bibinfo {editor} {\bibfnamefont {N.~J.}\ \bibnamefont {Salkind}},\ ed.,\
  \href@noop {} {\emph {\bibinfo {title} {Encyclopedia of measurement and
  statistics}}}\ (\bibinfo  {publisher} {SAGE Publications},\ \bibinfo {year}
  {2006})\BibitemShut {NoStop}%
\bibitem [{\citenamefont {Cleophas}\ and\ \citenamefont
  {Zwinderman}(2012)}]{sacs}%
  \BibitemOpen
  \bibfield  {author} {\bibinfo {author} {\bibfnamefont {T.~J.}\ \bibnamefont
  {Cleophas}}\ and\ \bibinfo {author} {\bibfnamefont {A.~H.}\ \bibnamefont
  {Zwinderman}},\ }\href@noop {} {\emph {\bibinfo {title} {Statistics applied
  to clinical studies}}}\ (\bibinfo  {publisher} {Springer},\ \bibinfo {year}
  {2012})\BibitemShut {NoStop}%
\bibitem [{\citenamefont {Cross}\ and\ \citenamefont
  {Plunkett}(2014)}]{phys-pharm}%
  \BibitemOpen
  \bibfield  {author} {\bibinfo {author} {\bibfnamefont {M.~E.}\ \bibnamefont
  {Cross}}\ and\ \bibinfo {author} {\bibfnamefont {E.~V.~E.}\ \bibnamefont
  {Plunkett}},\ }\href@noop {} {\emph {\bibinfo {title} {Physics, pharmacology
  and physiology for anaesthetists}}},\ \bibinfo {edition} {2nd}\ ed.\
  (\bibinfo  {publisher} {Cambridge University Press},\ \bibinfo {year}
  {2014})\BibitemShut {NoStop}%
\bibitem [{\citenamefont {Franklin}(2013)}]{standards}%
  \BibitemOpen
  \bibfield  {author} {\bibinfo {author} {\bibfnamefont {A.}~\bibnamefont
  {Franklin}},\ }\href@noop {} {\emph {\bibinfo {title} {Shifting standards:
  Experiments in particle physics in the twentieth century}}}\ (\bibinfo
  {publisher} {University of Pittsburgh Press},\ \bibinfo {year}
  {2013})\BibitemShut {NoStop}%
\bibitem [{\citenamefont {Carroll}(2013)}]{particle}%
  \BibitemOpen
  \bibfield  {author} {\bibinfo {author} {\bibfnamefont {S.}~\bibnamefont
  {Carroll}},\ }\href@noop {} {\emph {\bibinfo {title} {The particle at the end
  of the universe: How the hunt for the Higgs boson leads us to the edge of a
  new world}}}\ (\bibinfo  {publisher} {Plume},\ \bibinfo {year}
  {2013})\BibitemShut {NoStop}%
\bibitem [{\citenamefont {Abbott}(2016)}]{PhysRevLett.116.061102}%
  \BibitemOpen
  \bibfield  {author} {\bibinfo {author} {\bibfnamefont {B.~P.}\ \bibnamefont
  {Abbott}} (\bibinfo {collaboration} {LIGO Scientific Collaboration and Virgo
  Collaboration}),\ }\href@noop {} {\bibfield  {journal} {\bibinfo  {journal}
  {Phys. Rev. Lett.}\ }\textbf {\bibinfo {volume} {116}},\ \bibinfo {pages}
  {061102} (\bibinfo {year} {2016})}\BibitemShut {NoStop}%
\bibitem [{\citenamefont {Abe}\ \emph {et~al.}(1994)\citenamefont {Abe} \emph
  {et~al.}}]{PhysRevD.50.2966}%
  \BibitemOpen
  \bibfield  {author} {\bibinfo {author} {\bibfnamefont {F.}~\bibnamefont
  {Abe}} \emph {et~al.},\ }\href@noop {} {\bibfield  {journal} {\bibinfo
  {journal} {Phys. Rev. D}\ }\textbf {\bibinfo {volume} {50}},\ \bibinfo
  {pages} {2966} (\bibinfo {year} {1994})}\BibitemShut {NoStop}%
\bibitem [{\citenamefont {Abachi}\ \emph {et~al.}(1995)\citenamefont {Abachi}
  \emph {et~al.}}]{PhysRevLett.74.2632}%
  \BibitemOpen
  \bibfield  {author} {\bibinfo {author} {\bibfnamefont {S.}~\bibnamefont
  {Abachi}} \emph {et~al.} (\bibinfo {collaboration} {D0 Collaboration}),\
  }\href@noop {} {\bibfield  {journal} {\bibinfo  {journal} {Phys. Rev. Lett.}\
  }\textbf {\bibinfo {volume} {74}},\ \bibinfo {pages} {2632} (\bibinfo {year}
  {1995})}\BibitemShut {NoStop}%
\bibitem [{\citenamefont {Abe}\ \emph {et~al.}(1995)\citenamefont {Abe} \emph
  {et~al.}}]{PhysRevLett.74.2626}%
  \BibitemOpen
  \bibfield  {author} {\bibinfo {author} {\bibfnamefont {F.}~\bibnamefont
  {Abe}} \emph {et~al.} (\bibinfo {collaboration} {CDF Collaboration}),\
  }\href@noop {} {\bibfield  {journal} {\bibinfo  {journal} {Phys. Rev. Lett.}\
  }\textbf {\bibinfo {volume} {74}},\ \bibinfo {pages} {2626} (\bibinfo {year}
  {1995})}\BibitemShut {NoStop}%
\bibitem [{\citenamefont {van Loock}\ and\ \citenamefont
  {Furusawa}(2003)}]{PhysRevA.67.052315}%
  \BibitemOpen
  \bibfield  {author} {\bibinfo {author} {\bibfnamefont {P.}~\bibnamefont {van
  Loock}}\ and\ \bibinfo {author} {\bibfnamefont {A.}~\bibnamefont
  {Furusawa}},\ }\href@noop {} {\bibfield  {journal} {\bibinfo  {journal}
  {Phys. Rev. A}\ }\textbf {\bibinfo {volume} {67}},\ \bibinfo {pages} {052315}
  (\bibinfo {year} {2003})}\BibitemShut {NoStop}%
\bibitem [{\citenamefont {Gerke}\ \emph {et~al.}(2015)\citenamefont {Gerke},
  \citenamefont {Sperling}, \citenamefont {Vogel}, \citenamefont {Cai},
  \citenamefont {Roslund}, \citenamefont {Treps},\ and\ \citenamefont
  {Fabre}}]{PhysRevLett.114.050501}%
  \BibitemOpen
  \bibfield  {author} {\bibinfo {author} {\bibfnamefont {S.}~\bibnamefont
  {Gerke}}, \bibinfo {author} {\bibfnamefont {J.}~\bibnamefont {Sperling}},
  \bibinfo {author} {\bibfnamefont {W.}~\bibnamefont {Vogel}}, \bibinfo
  {author} {\bibfnamefont {Y.}~\bibnamefont {Cai}}, \bibinfo {author}
  {\bibfnamefont {J.}~\bibnamefont {Roslund}}, \bibinfo {author} {\bibfnamefont
  {N.}~\bibnamefont {Treps}}, \ and\ \bibinfo {author} {\bibfnamefont
  {C.}~\bibnamefont {Fabre}},\ }\href@noop {} {\bibfield  {journal} {\bibinfo
  {journal} {Phys. Rev. Lett.}\ }\textbf {\bibinfo {volume} {114}},\ \bibinfo
  {pages} {050501} (\bibinfo {year} {2015})}\BibitemShut {NoStop}%
\bibitem [{\citenamefont {Shchukin}\ and\ \citenamefont {van
  Loock}(2015)}]{PhysRevA.92.042328}%
  \BibitemOpen
  \bibfield  {author} {\bibinfo {author} {\bibfnamefont {E.}~\bibnamefont
  {Shchukin}}\ and\ \bibinfo {author} {\bibfnamefont {P.}~\bibnamefont {van
  Loock}},\ }\href@noop {} {\bibfield  {journal} {\bibinfo  {journal} {Phys.
  Rev. A}\ }\textbf {\bibinfo {volume} {92}},\ \bibinfo {pages} {042328}
  (\bibinfo {year} {2015})}\BibitemShut {NoStop}%
\bibitem [{Note1()}]{Note1}%
  \BibitemOpen
  \bibinfo {note} {\protect \texttt {http://cvxopt.org}}\BibitemShut {NoStop}%
\end{thebibliography}
\end{document}